
\input phyzzx
\input tables
\Pubnum{FERMILAB-Pub-93/105-T}
\date{May, 1993}
%
\overfullrule0pt
\def\refmark#1{[#1]}
\def\subsection#1{\par
   \ifnum\the\lastpenalty=30000\else \penalty-100\smallskip \fi
   \noindent{\enspace\it{{#1}}}\enspace \vadjust{\penalty5000}}

%
\def\mynot#1{\not{}{\mskip-3.5mu}#1 }

\def\sqr#1#2{{\vcenter{\vbox{\hrule height.#2pt
        \hbox{\vrule width.#2pt height#1pt \kern#1pt
           \vrule width.#2pt}
       \hrule height.#2pt}}}}

\def\cite#1{\Ref~#1 }


\newcount\lineno
{\catcode`\`=\active \gdef`{\relax\lq}}

{\obeyspaces\global\let =\ } 

\def\emu{e^\pm \mu^\mp}
\def\mt{m_t}

\def\tautau{\tau^+ \tau^-}
\def\ww{W^+W^-}
\def\to{\rightarrow}

\def\subsection#1{\par
   \ifnum\the\lastpenalty=30000\else \penalty-100\smallskip \fi
   \noindent{\enspace\it{{#1}}}\enspace \vadjust{\penalty5000}}

%
\titlepage
\doublespace
\title{\seventeenrm HEAVY TOP QUARK SEARCHES IN THE
DI-LEPTON MODE AT THE TEVATRON}
\vskip0.5in
\centerline{Tao Han and Stephen Parke}
\vskip10pt
\centerline{ \it Fermi National Accelerator Laboratory}
\centerline{\it P.O. Box 500, Batavia, IL  60510, U.S.A.}
\centerline{\it (HAN@FNAL) ~~ (PARKE@FNAL)}
\vskip0.5cm
\abstract{
We present the results of a detailed
study of the effects of $b$-tagging on the heavy top-quark signal and
backgrounds for the modes of the
di-lepton plus two high transverse energy jets
at the Fermilab Tevatron.
The  general characteristics of the heavy top-quark
signal events are also discussed so that a comparison
can be made between  $b$-tagging  and
imposing stringent kinematical cuts to eliminate backgrounds.}
\endpage


\REF\topnew{For a recent review, see, {\it e. g.}, G. Kane, in
Mexico City High Energy Phenomenology 1991, pg 241.}
%
\REF\paul{P. Langacker, Univ. of Pennsylvania Preprint,
UPR-0555-T Mar. 1993.}
\REF\topone{A.~Caner for the CDF Collaboration,
XXVIIIth Rencontres de Moriond, Mar. 1993, to be published.}
\REF\toptwo{M.~Narian for the D0 Collaboration,
XXVIIIth Rencontres de Moriond, Mar. 1993, to be published.}
\REF\kunszt{Z. Kunszt and W. J. Stirling, Phys. Rev. {\bf D37}, 2439 (1988).}
\REF\gielea{W. Giele, D. Kosower, and H. Kuijf, Nucl. Phys.
(Proc. Suppl.) {\bf 23B}, 22 (1991); J.~M.~Benlloch, N.~Wainer, and
W. Giele, FERMILAB-PUB-93/060-T.}
\REF\bargera{H. Baer, V. Barger, and R. J. N. Phillips, Phys. Rev. {\bf D39},
3310 (1989). }
\REF\gieleb{F. A. Berends, J. B. Tausk, and W. T. Giele,
Phys. Rev. {\bf D47}, 2746 (1993).}
\REF\gielec{F. A. Berends, H.~Kuijf, B. Tausk, and W.~T.~Giele, Nucl. Phys.
{\bf B357}, 32 (1991).}
\REF\topme{R. Kleiss and W. J. Stirling, Z. Phys. {\bf C40}, 419 (1988).}
\REF\dyme{V. Barger, T. Han, J. Ohnemus, and D. Zeppenfeld,
Phys. Rev. {\bf D40}, 2888 (1989).}
\REF\wwme{V. Barger, T. Han, J. Ohnemus, and D. Zeppenfeld,
Phys. Rev. {\bf D41}, 2782 (1990).}
\REF\ellis{R. K. Ellis, Phys. Lett. {\bf B259}, 492 (1991).}
\REF\hmrsb{ P. Harriman, A. Martin, R. Roberts, and W. J. Stirling,
Phys. Rev. {\bf D42}, 798  (1990).}
\REF\fnote{
For these processes, at Tevatron energies,
we consider the  $b$-flavor partons inside a proton
to contribute to higher order corrections to
$b\bar b$ production processes.
Our uncertainty estimates include the effects of
such higher order corrections.}
\REF\bargerb{H. Baer, V. Barger, J. Ohnemus, and R. J. N. Phillips,
Phys. Rev. {\bf D42}, 54 (1990).}

In spite of the tremendous success of the Standard Model (SM)
up to the highest energies accessible  today,
the top quark has not yet been  observed.
To experimentally establish
the existence of the top quark ($t$)
and measure  its mass ($m_t$) is not
only crucial to completing
the pattern of the three generations of fermion families,
but it is also important as a  probe of new physics beyond the
SM \refmark{\topnew}.

Precision measurements of electroweak parameters imply that
in the SM, \nextline
$m_t = 150^{+19+15}_{-24-20}$~GeV \refmark{\paul}.
The current direct experimental mass limits on the SM top quark are
$m_t>108$ GeV from CDF
Collaboration \refmark{\topone} and $m_t>103$ GeV from D0 Collaboration
\refmark{\toptwo}. With an integrated luminosity of 25 pb$^{-1}$ for
the Tevatron, CDF/D0 should be able to search for the top quark
up to $m_t \approx 130$ GeV;
and to $m_t \approx 170$ GeV with 100 pb$^{-1}$.
With the new Main Injector an integrated luminosity of
$10^3$ pb$^{-1}$ should be easily obtained, pushing
the reach of the Tevatron to top-quark masses greater than
$200$ GeV.

Standard Model top quarks, with masses  greater than the $W$-boson mass,
are  produced at  Tevatron
energies,  $\sqrt s = 1.8$ TeV, predominantly by
$q \bar q \rightarrow t \bar t$ and
decay into $Wb$ final state with almost a 100\% branching fraction.
The experimental signature of a top
quark will therefore be two energetic $b$-quark
jets plus a pair of $W$'s.
The hadronic decays of the $W$-boson pair are of the largest rate for the
top-quark signal, with a branching fraction
BF($t\bar t \rightarrow b \bar b + 4$ jets)=(2/3)$^2$. However, the QCD 6-jets
background  swamps this signal \refmark{\kunszt}. Even with an efficient
$b$-tagging, the experimental searches in this channel are still difficult
\refmark{\gielea}. The single leptonic modes also have a large rate, with
BF($t\bar t \to b \bar b \ell^\pm \nu + $ 2 jets)=2(2/9)(2/3), where
$l=e,\mu$. The continuum $W^\pm+$ 4-jets background is also
large.  For a top quark of $m_t < 170$ GeV, it is possible after some
optimized kinematical cuts to reach a
signal/background ratio of order
one \refmark{\bargera,\gieleb}.
Additional information can be obtained from $b$-tagging
to improve the situation \refmark{\gielec},
however more comprehensive studies
including the effects of $c$-quark production and jet
smearing are required.
The di-lepton modes
are considered to be the cleanest channel for top quark searches,
though with a smaller branching fraction
BF($t\bar t \to b \bar b \ell^+ \ell^- \nu {\bar \nu}$)=$(2/9)^2$.
For a very heavy top quark this channel
is not background free.

In this letter, we study the effects of $b$-tagging to purify the
top-quark signal in the di-lepton decay modes.
For simplicity, only $e^\pm \mu^\mp$ final states
are counted in the numerical presentation.
An additional factor of two is needed to
include $ee$ and $\mu \mu$ channels but  slightly more care is
necessary to separate the
backgrounds.
We present for the first time
the calculation of the background processes,
$p \bar p \to  \emu ~{\mynot p}_T ~b \bar b $,
$\emu ~{\mynot p}_T ~c \bar c , \
{\rm and} \ \emu ~{\mynot p}_T ~c q(g) $
(where ${\mynot p}_T$ denotes the missing transverse momentum),
which are relevant to the $b$-tagged  signal.
A comparison is made between the effects  of tagging one $b$-quark jet versus
imposing stringent kinematical cuts  to eliminate backgrounds.

The experimental signature of a heavy top quark in the di-lepton mode
under consideration is
$$
t\bar t\ \rightarrow\ \emu \ {\mynot p}_T j j\ ,
\eqn\ONE
$$
where $\emu$ are well isolated charged leptons
coming from the $W$ decays; ${\mynot p}_T$ is
mainly from the missing neutrinos; and $j$'s are the jets resulting from the
$b$-quark hadronization and decay.

Without identifying the $b$-quark flavor, the major backgrounds
to process ~\ONE~ are the Drell-Yan process
(called $\tau^+\tau^-jj$ background)
$$
 Z^*(\gamma^*) jj\ \rightarrow\ \tau^+ \tau^- jj \ \rightarrow\ \emu \
{\mynot p}_T j j\ ,
\eqn\TWO
$$
and the $W$-pair production (called $W^+W^-jj$ background),
$$
W^+ W^- jj\ \rightarrow\ \emu \ {\mynot p}_T j j\ . \eqn\THREE
$$

We have calculated the cross section for these three
processes using the fully
spin-correlated matrix elements \refmark{\topme-\wwme}.
For the production of a pair of top quarks,
heavier than 100 GeV at Tevatron energies, the typical
momentum fraction of a parton inside a proton is in the
range  0.1 -- 0.3.
Therefore,
the theoretical uncertainty from the
parton distribution functions is small.
However, since our calculations are
at tree level, ${\cal O}(\alpha_s^2(Q))$,
the results are sensitive to
our choice of the renormalization scale, $Q$.
We normalize our signal cross section to
the full ${\cal O}(\alpha_s^3)$
results \refmark{\ellis}, which has
a small uncertainty, approximately 25\%.
For the backgrounds, we estimate the uncertainties
by varying the scale in the range
$M_Z^{}/2 < Q < 2M_Z^{}$ for $\tau^+ \tau^-jj$
and $M_W^{}/2 < Q < M_{WW}^{}$ for $W^+ W^-jj$, where $M_{WW}^{}$ is
the invariant mass of the $W^+W^-$ system.
The HMRS Set-B parton distribution functions \refmark{\hmrsb} have
been used for these calculations with the
factorization scale set equal to the renormalization scale, $Q$.

To calculate the cross sections for both
the signal and background processes
we first impose the following minimal acceptance cuts,
$$\eqalign{
p_T^{}(\ell)\ >&\ 15\ {\rm GeV}\ , \qquad |y(\ell)|\ <\ 2\ ,
\qquad  {\mynot p}_T\ > \ 20 {\rm GeV}\ , \cr
E_T^{}(j) >&\ 15\ {\rm GeV}\ , \qquad |y(j)|\ <\ 2\ , \cr
\Delta R(jj) >&\ 0.7\ , \qquad  \qquad \Delta R(\ell j, {\rm or} \
\ell \ell)\ >\ 0.4\ , \cr}
\eqn\FOUR
$$
where $p_T^{}(E_T^{})$, $y$, and $\Delta R(ij)$
are the transverse momentum (energy),
pseudorapidity, and the separation between $i$ and $j$ in the
pseudorapidity-azimuthal angle plane, respectively.
These cuts cover most of the phase space region for the signal events
and roughly simulate the CDF/D0 detector acceptance.

\FIG\FONE{Total cross sections
at $\sqrt s = 1.8$ TeV, with the minimal cuts, Eq. {\FOUR},
for the $t \bar t$ signal (dashes), and backgrounds from
 $\tau^+ \tau^- jj$ and $W^+ W^- jj$ (dots).
The solid lines are for the processes,
 $\tau^+ \tau^- jj$ and $W^+ W^- jj$,
with at least one of the jets a $b$-quark or $c$-quark jet.
The range corresponds
to the scale choices of $M_Z^{}/2 < Q < 2M_Z^{}$ for
$\tau^+ \tau^-jj$ and
$M_W^{}/2 < Q < M_{WW}^{}$ for $W^+ W^-jj$.}

Figure {\FONE} shows the total cross sections for the
processes {\ONE-\THREE}.
For $m_t < 120$ GeV, we see  that the signal
rate (dashes) is larger than the backgrounds (dots)
by about an order of magnitude.
The total background cross section is  less than 46 fb (38 fb from
$\tau^+\tau^-jj$ channel), corresponding
to 1.1 $e \mu$ events for an integrated luminosity of 25 pb$^{-1}$.
In  searching for a heavier top quark additional handles to
suppress these backgrounds are desirable.

Since CDF has a working Silicon microvertex detector,
it is natural to ask how advantageous it is to tag on one of
the $b$-quarks to enhance the top-quark signal over the background.
If the jets in
background processes were all from light quarks (or gluons), then the
backgrounds would be highly suppressed by tagging one of the
two $b$-quarks in the signal,
assuming a 1\% misidentification for the light quark jets.
However, there is open flavor $b \bar b$ production via
$$\eqalign{
gg, \  q \bar q & \to Z^*(\gamma^*) b \bar b \
\to \ \tau^+ \tau^- b \bar b
\to \ \emu \ {\mynot p}_T b \bar b\ , \cr
gg, \ q \bar q & \to W^+ W^- b \bar b \
\to\ \emu \ {\mynot p}_T b \bar b \ .\cr}
\eqn\THREEP
$$
which have to be carefully studied when $b$-tagging is applied.
We have calculated these cross sections using the full Standard Model
matrix elements, in the limit $m_b = 0$.
The contribution from the  $b$-parton distribution in
the protons has been ignored \refmark{\fnote}.
The $b$-quark mixings with the first two generation quarks
are small ({\it e. g.}, $V_{cb}^2 \approx 3 \times 10^{-3}$)
and hence negligible.
With the cuts of
Eq.~\FOUR, we find that the cross section for $\emu {\mynot p}_T b \bar b$
is 0.21 fb from $W^+W^-jj$ process  and  0.47 fb from $\tau^+ \tau^-jj$.
These background rates are significantly
smaller than the signal, making the $b$-tagging scheme very promising
for enhancing the signal to background ratio.

However, due to the comparable lifetime of the
charmed-mesons to the $b$-mesons, it is difficult to experimentally
distinguish the $b$'s and $c$'s on an event by event basis. We are
therefore forced to consider the $c$-quark production.
Besides the open flavor $c \bar c$ pair production similar
to Eq.~\THREEP,
there are single $c$ contributions, such as
$$
g c \to \emu {\mynot p}_T g c, \qquad \ qc, \ qq' \to \emu {\mynot p}_T q c,\
\eqn\FOURP
$$
that must be evaluated if  single $b$-tagging is to be used.
Again, we have ignored the generation mixings among quarks. The largest
error in this approximation is from the valence quark
transition $d \to cW$, proportional to $V_{cd}^2=0.048$.
The cross section from this transition is only about 10$^{-3}$
fb, which is negligible.

In Table~\ONE~ we list the contributions from
the individual channels.
Figure~1 compares the cross sections of
$\emu {\mynot p}_T jj$ with at least one heavy quark ($b, \ c$)
in the jets from the $t \bar t$ signal, $W^+W^-jj$ and $\tau^+ \tau^-jj$
background processes.
{}From this figure it is clear that the signal rate is higher than the
backgrounds by more than an order of magnitude up to
$m_t \approx 200$~GeV.
Assuming a 30\% tagging efficiency, independent of $E_T$,
for a $b$-quark to be identified
having a second vertex (50\% for tagging one of the two $b$'s),
the signal cross section for $m_t \approx 200$~GeV
is  16 fb, while the summed background is less than 1.0 fb.
Now, the signal to background ratio is
better than 16:1.
To achieve this high S/B ratio,
the signal has been reduced by a factor of two.
However,
the experimentalists
could  loosen  the $b$-tagging
requirement to retain a larger fraction of the signal,
since the background is so much smaller than the signal.
Table~\ONE~ can be used to calculate the backgrounds
for tagging efficiencies other than the 30\% assumed here.
Of course in reality the $b$-tagging efficiency is
higher for more energetic jets.
Since the signal jets are typically stiffer than the background jets,
the signal to background ratio will be higher than indicated here.

\TABLE\hqtable{}
\topinsert
\titlestyle{\twelvepoint
Table~\hqtable: The central value for the
QCD $b$-quark and $c$-quark associated production rates
with  $\emu$   and ${\mynot p}_T$ at $\sqrt s=1.8$  TeV
(in units of fb). $W^+W^-$ and $\tau^+\tau^-$ indicate the source
of the final state $\emu$.}
\bigskip
\thicksize=0pt
\hrule \vskip .04in \hrule
\begintable
%
%
$ {} $ & $~~W^+ W^-~~$ & $~~\tau^+\tau^-~~$ \cr
$ q \bar q \to \emu {\mynot p}_T b \bar b $ & 0.20  & 0.33 \nr
$ gg \to \emu {\mynot p}_T b \bar b $       & 0.01  & 0.14 \cr
$ q \bar q \to \emu {\mynot p}_T c \bar c $ & 0.10  & 0.30 \nr
$ gg \to \emu {\mynot p}_T c \bar c $       & 0.01  & 0.11 \nr
$ gc \to \emu {\mynot p}_T  g c $           & 0.02  & 0.30 \nr
$ qc(q') \to \emu {\mynot p}_T q c $        & 0.01  & 0.15 \endtable

\vskip .04in
\hrule \vskip .04in \hrule
\endinsert

\FIG\FTHREE{Differential cross sections, at $\sqrt s = $ 1.8 TeV,
for the signal (solid) with
$m_t=150$ GeV and 200 GeV and the backgrounds $\tautau jj$ (dot-dashed)
and $\ww jj$ (dashed); (a). Minimum $p_T^{}$ distribution of the two charged
leptons, (b). Leptonic cluster transverse mass, $M_T$, distribution.
The minimal cuts of Eq.~\FOUR~ have been imposed.}

\FIG\FFOUR{Differential cross sections, at $\sqrt s = $ 1.8 TeV,
for the signal (solid) with
$m_t=150$ GeV and 200 GeV and the backgrounds $\tautau jj$ (dot-dashed)
and $\ww jj$ (dashed); (a). Minimum $E_T$ distribution of the two jets,
(b). The scalar sum of the transverse jet energies,
$E_T^{sum}$, distribution.
The minimal cuts of Eq.~\FOUR~ have been imposed.}

\FIG\FTWO{Total cross sections
at $\sqrt s = 1.8$ TeV, with stringent cuts, Eq. {\FOUR} and {\FIVEN},
for the $t \bar t$ signal (dashes), and backgrounds from
 $\tau^+ \tau^- jj$ and $W^+ W^- jj$ (dots).
The range corresponds
to the same choices of scale as Fig. {\FONE}.}

An alternative technique  for separating the di-lepton
signal from the backgrounds
is to tighten the kinematical cuts
\refmark{\bargerb}.
Since the $\tautau jj$ background is predominantly
from an on-shell $Z$ decay,
the cluster transverse mass variable $M_T$ is  limited by the
$Z$ mass, where $M_T$ is defined as
$$
M_T^2=[(M^2_{\ell\ell}+ p^2_{T\ell\ell})^{1/2} + |{\mynot p}_T|]^2
- ( {\bf p}_{T\ell\ell} + {\mynot {\bf p}}_T )^2
\eqn\FIVE
$$
with $M_{\ell\ell}$ and $p_{T\ell\ell}$ the the invariant mass and the
transverse momentum of the charged lepton pair.
As suggested in Ref.~\bargerb,
requiring $M_T > M_Z$ would therefore substantially reduce
the $\tautau$ background.
In Figure {\FTHREE}(a) and (b) we have plotted
the signal and background differential cross sections
versus the minimum transverse momentum
of the charged leptons, $p_T^{min}(l)= min(p_T(l_1),p_T(l_2))$,
and the leptonic cluster transverse mass, $M_T$.
For the $\emu$ modes with two high $E_T$ jets, the distributions
of the azimuthal separation between the two
charged leptons does not provide a clear
distinction between the signal and backgrounds.
In particular the high transverse energy of the jets
smears out the peak in the
 $\tau^+ \tau^- jj$ background at 180 degrees.

Furthermore, the QCD jets in the backgrounds
discussed here tend to be soft, in contrast to the rather hard $b$-quark
jets from heavy top decays.
A higher jet threshold would be helpful in reducing the backgrounds.
To see this, in Figure {\FFOUR}(a) and (b) we have plotted
the signal and background differential cross sections
versus the minimum transverse energy
of the two jets, $E_T^{min}(j)= min(E_T(j_1),E_T(j_2))$,
and the scalar sum of the transverse jet energies, $E_T^{sum} ~=~
E_T(j_1) + E_T(j_2)$.

Figure {\FTWO} shows the total cross section with the additional cuts
$$\eqalign{
M_T^{}\ >\ 90\ {\rm GeV}, \qquad \  p_T^{}(j) >\ 30\ {\rm GeV}. \cr}
\eqn\FIVEN
$$
With these stringent cuts, Eq. {\FOUR} and {\FIVEN},
the signal rate is  larger than the backgrounds by
an order of magnitude up to $\mt = 200$ GeV,
with only a moderate reduction in the signal;
about 30\% reduction for $m_t=150$~GeV, and 10\% for $m_t=200$~GeV.
Further kinematic cuts could also be applied, if necessary,
such as a cut on the scalar sum of the transverse
jet energies.

In conclusion, we have demonstrated that in the
search for the heavy top quark in the di-lepton
mode that the signal can be easily distinguished from the
$\ww jj$ and $\tautau jj$ physics backgrounds by
either tagging one of the jets as a $b$-quark jet or
by making stringent kinematical cuts for top quark
masses up to 200 GeV.
Of course, events that are $b$-tagged and also
pass the stringent kinematical cuts are truly
platinum candidates for the discovery of the heavy top quark.

\bigskip
\bigskip
\bigskip

\ack

The authors wish to thank V.~Barger, W.~Giele, J.~Ohnemus, M.~Peskin,
R.~Phillips, S.~Willenbrock and G.-P.~Yeh for discussions.
Fermilab is operated by the Universities Research Association
Inc. under contract with the United States Department of Energy.

\endpage
\refout
\endpage
\figout
\end